# The EU AI Act Beyond Europe:
## A Practical Readiness Framework for Non-EU Organisations. Comparative Evidence from Four African Countries

**Uloma Okoro**
Aula Fellowship for AI, Montreal, Canada
uloma@theaulafellowship.org
ORCID: 0009-0000-7720-8510

**Branislav Radeljić**
Aula Fellowship for AI, Montreal, Canada
branislav@theaulafellowship.org
ORCID: 0000-0002-0497-3470

**Abstract**
The EU Artificial Intelligence Act has implications for organizations within Europe as well as organisations outside the European Union where they provide AI-enabled services to EU clients, use European vendors, participate in EU-facing supply chains, or generate outputs that are used in the EU. This study examines the implications of these impacts for African countries and other non-EU organisations, using South Africa, Nigeria, Kenya and Egypt as comparative cases. It applies doctrinal legal analysis, structured document review and scenario-based comparison to assess the sufficiency or otherwise of existing regulatory frameworks within the selected countries. The analysis shows that the existing regulatory frameworks (like data-protection, financial-sector, cybersecurity and corporate-governance rules) can support AI readiness, but they do not resolve questions of territorial scope, role classification, risk categorisation, vendor responsibility and access to compliance evidence. The analysis proposes a seven-step readiness framework which covers exposure mapping, AI inventories, role and risk classification, governance, contracts, evidence management and continuous monitoring. It argues for proportionate, interoperable approaches that enable non-EU organisations to demonstrate readiness without treating EU compliance as a uniform exercise.

**Contents**

## Executive Summary

The EU Artificial Intelligence Act (2023) has entered a decisive stage of implementation. From 2 August 2026, most of its provisions became applicable, and this includes important requirements governing high-risk AI systems, transparency, regulatory supervision and enforcement.[1] Although the Act is an EU regulation, its effects are not confined to organisations established in Europe. African and other non-EU organisations may be affected where they supply AI systems or AI-enabled services to EU clients, produce outputs that are used in the EU, participate in European supply chains, use European AI vendors or operate within multinational groups subject to EU requirements.

This policy study sheds light on what those organisations should do in practice. It focuses on banks, fintechs, technology companies and other regulated organisations, using South Africa, Nigeria, Kenya and Egypt as comparative cases. These were selected because they offer different financial-market structures, regulatory institutions, legal traditions and levels of digital development. They are not presented as representative of Africa as a whole. Instead, they provide contrasting settings through which common regulatory and organisational problems can be tested. The study distinguishes four forms of exposure: (i) direct legal applicability under the AI Act; (ii) contractual exposure arising through customer agreements, procurement requirements, audit clauses and supplier obligations; (iii) market-access exposure, where providing evidence of AI governance becomes a practical condition for obtaining or retaining EU business; and (iv) supervisory or reputational exposure, particularly for regulated organisations whose use of AI may attract scrutiny even where the Act's territorial application is uncertain. The analysis shows that a company's exposure to the provisions of the Act depends on its role in the AI value chain, the purpose of the system, where outputs are used, who controls the system and which party holds the information required to demonstrate compliance. For instance, a credit score produced in Africa and relied upon by an EU lender may create a different legal and operational position from a fraud-detection tool used only within an African bank. Similarly, an organisation may acquire provider responsibilities if it substantially modifies or rebrands a system.[17] South Africa, Nigeria, Kenya and Egypt already possess relevant foundations in data protection, financial regulation, cybersecurity, outsourcing and corporate governance. These controls can support AI readiness, but they do not automatically answer AI Act-specific questions concerning role classification, high-risk systems, transparency obligations or territorial reach. The principal challenge is therefore often fragmentation rather than complete regulatory absence.

The study proposes a seven-step readiness framework for affected organisations: (i) map EU exposure; (ii) create an AI inventory; (iii) classify organisational roles and system risk; (iv) establish governance; (v) strengthen contracts and vendor controls; (vi) build and retain evidence; and (vii) monitor systems throughout their lifecycle. A maturity scale allows organisations to assess progress without treating readiness as legal certification. The study recommends dedicated third-country guidance from the European AI Office, coordinated sectoral guidance from African regulators, proportionate supplier requirements from EU clients, dual-track readiness programmes within non-EU organisations, and shared implementation resources developed by business associations. The study concludes that non-EU organisations should not wait for a regulator, customer or contract to define their exposure. They should identify it, document it and manage it as a matter of urgency.

## 1. The current policy problem

The EU Artificial Intelligence Act is no longer a distant compliance issue for organisations outside Europe. Its phased implementation has reached the point at which many of its central requirements are applicable, while EU customers, multinational groups and technology vendors are increasingly translating those requirements into procurement, contracting and governance expectations. For African banks, fintechs and technology companies, the immediate policy problem is therefore no longer simply whether the Act applies to them but whether they can identify and manage the different ways in which it may affect their business.

### *1.1 The regulatory position after August 2026*

The AI Act entered into force on 1 August 2024, but its requirements were introduced in stages. Prohibitions on certain AI practices and the obligation to promote sufficient AI literacy began to apply on 2 February 2025. Governance provisions and obligations relating to general-purpose AI models followed on 2 August 2025. From 2 August 2026, most of the Act became applicable, including important requirements governing transparency, high-risk AI systems, regulatory supervision and enforcement. Among the provisions taking effect on 2 August 2026 were the transparency obligations in Article 50 of the Act. These require, among other things, that people are informed when they are interacting directly with certain AI systems, unless this is already obvious from the circumstances. They also introduce disclosure or technical-marking requirements for specified AI-generated or manipulated content, including deepfakes, and require notification where emotion-recognition or biometric-categorisation systems are used. These obligations are relevant to non-EU organisations offering chatbots, biometric tools or content-generation services to EU clients, even where the systems concerned are not classified as high-risk.[1][2]

The timetable does not end there. Certain high-risk AI systems connected to products regulated under existing EU product-safety legislation benefit from a later transition period, generally extending to 2 August 2027. Providers of general-purpose AI models placed on the market before 2 August 2025 also have a later compliance deadline.[1]

This staggered implementation creates a difficult environment for non-EU organisations. The Regulation itself is legally binding, while European Commission guidelines and European AI Office interpretations assist organisations in understanding how it may be applied. The General-Purpose AI Code of Practice provides a voluntary compliance route for relevant model providers, but it does not replace the Regulation.[3][4] Harmonised technical standards may eventually help organisations demonstrate conformity, although their development and formal recognition follow a separate process.[5]

Several questions will continue to require careful interpretation. These include what it means for an AI-generated output to be "used" in the EU, when the modification of a third-party system is substantial enough to change an organisation's legal role, and how responsibility should be divided between providers, deployers, integrators and subcontractors.

### *1.2 How will non-EU organisations be affected?*

There are three principal exposure pathways that can help to answer this question.

The first is direct legal exposure. The Act can apply to a provider outside the EU where it places an AI system or general-purpose AI model on the EU market. It can also apply to certain non-EU providers and deployers where the output produced by an AI system is used within the EU.[1] For instance, a credit score generated in Lagos, Nairobi, Johannesburg or Cairo and relied upon by a European lender may therefore raise questions under the Act even where the system itself is operated outside Europe. Legal roles can also change. An organisation that places its name or trademark on another provider's system, substantially modifies the system, or changes its intended purpose may acquire provider responsibilities. Importers, distributors and authorised representatives may carry additional obligations within the EU value chain.

The second pathway is indirect contractual exposure. An African organisation may fall outside the clearest cases of territorial application but still face AI Act-related demands from an EU customer or multinational parent. These may appear in procurement questionnaires, supplier codes, warranties, audit clauses, outsourcing agreements, incident-notification requirements and contractual indemnities. In practice, access to the European market may depend on the organisation's ability to provide evidence about data quality, testing, human oversight, cybersecurity and model performance.

The third pathway arises where an African organisation uses a European AI vendor. The customer must understand the system's intended purpose, limitations, classification, update process and documentation. It should not rely unquestioningly on the vendor's classification because the way the system is deployed, integrated or modified may affect its legal treatment. This is commercially significant. For instance, a 2024 survey by the Bank of England and Financial Conduct Authority found that 75 per cent of responding financial firms were already using AI, while about one-third of reported AI use cases involved third-party implementation.[6] The Financial Stability Board has similarly identified third-party concentration and limited visibility into external models as emerging financial-sector risks.[7]

### *1.3 Why do regulated organisations require particular attention?*

Banks, fintechs and technology companies increasingly use AI for credit assessment, fraud detection, identity verification, customer profiling, recruitment, cybersecurity, transaction monitoring and customer-service chatbots. These uses do not all receive the same treatment. For instance, AI systems used to assess creditworthiness or determine access to essential services may be high-risk. Recruitment and some employee-management systems may also be high-risk, as may specified biometric applications. AI systems used for risk assessment and pricing in life and health insurance is specifically addressed by the Act.[1] By contrast, fraud detection, anti-money-laundering monitoring, cybersecurity tools and internal general-purpose AI applications are not automatically high-risk merely because they are used by a financial institution. They may nevertheless attract transparency, privacy, consumer-protection, model-risk, cybersecurity and contractual requirements. Chatbots may also require users to be informed that they are interacting with AI. Thus, the practical lesson in these scenarios is that regulatory significance cannot be determined from the label given by the technology manufacturers alone. A system that is not classified as high-risk under the AI Act may still create serious supervisory, consumer or operational consequences. Regulated organisations therefore need an integrated assessment that considers the EU AI Act alongside financial regulation, data protection, cybersecurity, outsourcing and corporate governance.[8]

### *1.4 The Selection of South Africa, Nigeria, Kenya and Egypt as Case Studies*

South Africa, Nigeria, Kenya and Egypt were selected because they provide contrasting but commercially relevant settings in which to examine the effects of the EU AI Act outside Europe, with particular focus on Africa. Together, they reflect differences in financial-market structure, regulatory maturity, legal tradition, technology adoption and relationships with European businesses. They also have sizeable banking, fintech or technology sectors in which AI is increasingly used for customer identification, credit assessment, fraud monitoring, cybersecurity and business automation. The comparison is therefore designed to show how the same EU-facing regulatory problem may require different organisational responses. Table 1 shows the relevance of the chosen countries to this study.

| Country | Comparative value |
|---|---|
| **South Africa** | South Africa provides the perspective of a comparatively developed banking and corporate-governance environment, with significant multinational activity and established institutions for financial supervision, data protection and corporate compliance. It is particularly useful for examining how the AI Act may interact with existing risk-management, outsourcing and accountability structures. |
| **Nigeria** | Nigeria represents a large and fast-moving fintech and technology market in which companies increasingly provide digital services across borders. Its developing AI and data-governance environment make it useful for considering how innovative firms can respond to EU requirements where technical capability may develop faster than formal compliance systems.[9] |
| **Kenya** | Kenya offers extensive experience in mobile financial services, digital credit and technology-enabled service delivery. Its active data-protection and financial-sector institutions provide a useful setting for examining how existing digital-finance controls can support AI readiness.[10] |
| **Egypt** | Egypt adds a North African and EU-facing perspective. Its civil-law tradition, national AI-policy programme, sizeable banking sector and established technology and business-process outsourcing industries make it relevant to questions of cross-border service provision, vendor relationships and regulatory alignment.[11] |

**Table 1: Countries chosen as case study and their relevance.**

The four countries are not intended to represent Africa as a whole, nor does their inclusion imply that their regulatory systems are more important than those of other African states. Africa contains substantial differences in market size, institutional capacity, legal systems, digital infrastructure and commercial relationships. The purpose of the four cases is to reveal different types of readiness challenge that may also arise elsewhere.

The comparison will avoid presenting each country as a separate descriptive case. Instead, the same business scenarios and analytical questions will be applied across all four. For example, the study will examine how an AI-enabled credit-scoring provider serving an EU bank might be affected if it were operating from Johannesburg, Lagos, Nairobi or Cairo. It will then consider whether differences in local financial regulation, data protection, vendor oversight and organisational capacity would alter the practical response. This approach makes the compareson useful to organisations rather than turning it into a ranking of national regulatory systems.

### *1.5 Methodology*

The study used a desk-based comparative legal and policy methodology. It began with a doctrinal analysis of the EU AI Act and relevant implementing materials to identify the legal rules, organisational roles and cross-border connections that may affect non-EU organisations. Primary legal and regulatory sources were distinguished from guidance, voluntary codes, expected standards and questions that remain unresolved.

This legal analysis was supported by structured document analysis. The evidence base includes legislation, regulatory guidance, parliamentary materials, official policy statements, national AI strategies, consultation documents, academic studies, institutional reports, corporate disclosures and reliable news reporting. Documents were not treated as equally authoritative. Greater weight was given to binding legal materials and official regulatory sources, while industry reports and news features were used mainly to illustrate market practices, implementation problems and emerging business responses. Document selection and review followed a systematic process. Each source was assessed for its authenticity, credibility, relevance, representativeness and meaning. This reduced the risk of treating frequently repeated online commentary as established evidence. It also reflects Bowen's view that document analysis should involve the careful selection, appraisal and interpretation of documents rather than their passive collection.[12] The four countries were examined through purposive comparative case selection. Cases were chosen because they expose relevant differences, not because they form a statistically representative sample. Purposive case selection is appropriate where the objective is to compare contrasting institutional settings and identify conditions that may explain different policy or organisational responses.[13]

The principal unit of analysis is the business scenario. Common scenarios include a bank using a European AI vendor, a technology company participating in an EU supply chain, and a multinational group operating a shared AI system across European and African entities. Applying the same scenarios across the four countries allowed the comparison to remain focused and consistent. The analysis also used regulatory analogy, particularly lessons from the contractual effects of the General Data Protection Regulation, financial-sector outsourcing rules and other EU regulatory regimes. These comparisons were used to identify likely organisational responses.

Important findings were triangulated wherever possible through three forms of evidence: a primary legal or official source, independent academic or policy analysis, and evidence of implementation or market practice. The resulting findings were then translated into an organisational readiness assessment, covering exposure mapping, role classification, governance, documentation, vendor management, contracting, monitoring and incident response. This method is intended to move the analysis from legal interpretation to practical action. Its central question is not merely whether the EU AI Act applies, but what a reasonable non-EU organisation should be able to identify, explain, document and demonstrate.

## 2. Why existing approaches are insufficient

Currently, the Existing guidance for the implementation of the EU AI Act provides an important foundation for understanding the Act. However, it is important to consider whether it fully resolves the practical questions facing organisations operating outside the EU. In this instance, it is necessary to consider whether there is a single method that connects legal scope, contractual relationships, local regulatory obligations and organisational capacity in such a manner as to ensure seamless compliance for non-EU countries.

### *2.1 Existing guidance remains predominantly EU-centred*

The European Commission guidance generally explains the Act through its legal categories: providers, deployers, importers, distributors, authorised representatives and providers of general-purpose AI models.[14] This is necessary, but then how do we factor in the operating conditions of a non-EU supplier or customer?

Some compliance materials appear to assume that an organisation already knows which systems qualify as AI, maintains an AI inventory and can access the documentation needed to classify its systems. They may also assume access to specialist legal, technical and assurance functions. These conditions are more likely to exist in a multinational organization than in a smaller financial institution, outsourced technology provider or local subsidiary using a system selected and controlled by its parent company. A non-EU customer may also be unable to obtain complete information from a vendor because of intellectual-property restrictions, security concerns or limited contractual rights. Guidance that begins with the applicable obligation may therefore be difficult to use where the organisation has not first established what systems it uses, who controls them and which party holds the relevant evidence.

### *2.2 Formal territorial scope does not necessarily capture the full practical effect*

Asking whether the Act legally applies remains essential, but it does not capture every form of exposure. Four distinct questions arise in this case. Legal applicability concerns whether the organisation falls within the Act's territorial and material scope. Contractual exposure arises where an EU client, provider or multinational group transfers information, monitoring or assurance requirements through an agreement. Market-access exposure concerns whether satisfying those requirements becomes a practical condition for obtaining or retaining EU business. Reputational or supervisory exposure may arise where an organisation's use of AI attracts concern from customers, investors, regulators or group risk functions, even where the precise application of the Act remains uncertain.

These categories are not legally equivalent. A contractual clause does not, by itself, make an organisation subject to every statutory obligation. However, EU providers and deployers may need information and cooperation from suppliers to meet their own responsibilities, particularly across complex AI value chains. Research on the external influence of EU digital regulation supports examining these commercial and institutional channels, while also cautioning that EU rules do not produce uniform effects in every foreign market.[15]

### *2.3 Existing African approaches may be fragmented*

The relevant starting point in Africa is not an absence of existing governance policies and control measure. South Africa, Nigeria, Kenya and Egypt already regulate aspects of AI-related activity through combinations of data-protection law, cybersecurity requirements, financial-sector supervision, consumer-protection rules, outsourcing controls and corporate-governance duties. National AI strategies and the African Union Continental Artificial Intelligence Strategy add broader policy objectives concerning responsible innovation, institutional capacity and continental cooperation.[16]

The practical difficulty is that these controls are distributed across different laws and regulators. A bank may assess privacy risk under data-protection law, operational risk under banking rules and cybersecurity risk under a separate framework, without having a single process for determining its role under the EU AI Act or establishing whether an output is used in Europe. Existing local controls may support readiness, but they do not automatically answer EU-specific questions.

### *2.4 Compliance materials often begin too late*

A checklist that begins by listing legal obligations may assume that the organisation has already completed the most difficult part of the exercise. Before applying any obligation, it must determine the following issues:

- which AI systems it uses or supplies;
- the role it performs in relation to each system;
- where the system and its outputs are used;
- which party controls its purpose and operation;
- whether it has been modified or rebranded; and
- who possesses the documentation required to support compliance.

These questions reflect the structure of the Act, under which obligations depend on the organisation's role, the system's purpose and classification, and changes made across the value chain.[1] Without this preliminary mapping, an organisation may overlook a relevant system or spend resources applying controls that do not correspond to its actual exposure.

### *2.5 Smaller organisations face a proportionality problem*

Organisations do not begin from the same position. A multinational bank may have established legal, model-risk, cybersecurity and procurement functions. A large fintech may possess advanced technical capability but less formal documentation. A medium-sized supplier may depend on customer specifications, while a start-up using third-party models may have limited access to information about model development and testing.

A practical framework must therefore establish a common minimum baseline while reserving enhanced controls for higher-risk uses, complex value chains and organisations with greater control over system design or deployment. This approach reflects the Act's own recognition that implementation support and compliance measures should take account of the needs of smaller enterprises.[1]

## 3. Analysis

### *3.1 Analytical framework*

The analysis applies the same sequence of questions to each country and business scenario. It first identifies the organisation's role in the AI value chain, because the EU AI Act assigns different obligations to providers, deployers, importers, distributors and other actors. It then examines the connection between the system or its output and the EU, the purpose for which the AI is used, and whether that use may fall within a high-risk or transparency category.

The analysis also distinguishes direct legal obligations from requirements transmitted through contracts, procurement processes and multinational group policies. Particular attention is paid to control of information. An organisation cannot demonstrate compliance merely by adopting a policy if another entity controls the model, training data, technical documentation, system updates or performance evidence.

The final questions are therefore practical and relate to which local regulatory controls can support readiness, where do gaps remain, and which party is capable of closing them?

### *3.2 Comparative regulatory context*

The four countries used as case study have established AI governance frameworks suited to their own systems for the time being. Each has data-protection, financial-sector, cybersecurity and consumer-protection rules that may already require elements of governance, risk assessment, security and accountability. Each has also adopted or developed a national AI-policy framework. These measures can provide a foundation for readiness, but none automatically determines whether an organisation is a provider or deployer under the EU AI Act, whether an output is used in the EU, or whether a particular application is high-risk.

South Africa's National AI Policy Framework sets out policy objectives for the development and governance of AI, while the Protection of Personal Information Act regulates the processing of personal information.[18][19] Financial institutions operate within an established supervisory structure involving the Prudential Authority and Financial Sector Conduct Authority, and the Intergovernmental Fintech Working Group has operated a regulatory sandbox. These arrangements may support risk management and regulatory engagement, but they do not constitute an AI Act conformity system.

Nigeria has developed a National Artificial Intelligence Strategy and enacted the Nigeria Data Protection Act 2023, supervised by the Nigeria Data Protection Commission.[20][21] The Central Bank of Nigeria regulates banks and payment-service providers and has issued a regulatory sandbox framework for financial innovation. The principal readiness issue is therefore not the total absence of relevant institutions, but the need to connect AI-specific questions with existing privacy, financial, cybersecurity and consumer-protection responsibilities.

Kenya's National Artificial Intelligence Strategy 2025–2030 sits alongside the Data Protection Act 2019 and sectoral oversight by the Central Bank of Kenya, the Office of the Data Protection Commissioner and the Capital Markets Authority.[22][23] Rules governing digital credit and existing innovation mechanisms make Kenya particularly relevant to AI-enabled lending and financial services. However, compliance with domestic digital-credit or privacy rules does not by itself establish the role classification, technical documentation or post-market monitoring required under the EU framework.

Egypt's second National Artificial Intelligence Strategy covers the period 2025–2030. The country also has a statutory personal-data framework and financial-sector oversight led by the Central Bank of Egypt, including a fintech regulatory

sandbox.[24][25] Egypt's banking, technology and outsourced-service sectors provide plausible pathways through which organisations may supply services to European clients or use systems obtained from European vendors.

The African Union Continental Artificial Intelligence Strategy provides an additional regional policy layer, including objectives relating to responsible AI, institutional capacity, data governance and international cooperation.[26] It is a strategic framework rather than a substitute for national law or EU compliance analysis. Table 2 shows the existing regulatory frameworks for the governance of Ai in the four selected countries. It should be noted that Table 2 is not intended to rank regulatory systems. Its purpose is to show that the same EU obligation may be supported by different domestic institutions and may encounter different implementation constraints.

| Issue | South Africa | Nigeria | Kenya | Egypt |
|---|---|---|---|---|
| **AI-policy framework** | National policy framework | National AI strategy | National AI Strategy 2025–2030 | National AI Strategy 2025–2030 |
| **Data-protection framework** | POPIA and Information Regulator | Data Protection Act 2023 and NDPC | Data Protection Act 2019 and ODPC | Personal Data Protection Law No. 151 of 2020 |
| **Financial-sector relevance** | Banks, insurers, fintechs and multinational groups | Large banking, payment and fintech market | Digital credit, mobile finance and capital-markets innovation | Banking, fintech and technology-enabled services |
| **Vendor-risk foundation** | Financial governance, outsourcing and information-security controls | Central-bank and sectoral risk requirements | Banking, digital-credit and data-protection controls | Central-bank, data-protection and cybersecurity controls |
| **EU-facing pathway examined** | Multinational groups and service providers | Cross-border fintech and technology services | Digital-service and technology suppliers | Banking, outsourcing and technology services |
| **Principal readiness issue** | Connecting established governance to AI-specific classification | Coordinating rapidly developing AI policy with sectoral controls | Translating digital-finance controls into AI evidence | Coordinating AI strategy, financial regulation and cross-border contracting |

**Table 2: Existing regulatory frameworks for the governance of AI.**

### *3.3 Likely scenarios for the application of AI systems and the effect of the EU AI Act*

*Scenario One: an African fintech supplying an EU client*

Consider a fintech that supplies credit-scoring, identity-verification, fraud-detection or transaction-monitoring services to an EU bank. If the fintech develops the system, or has it developed and supplies it under its own name or trademark, it may qualify as a provider. Supplying the system for use by an EU customer may also amount to placing it on the EU market, depending on the contractual and technical arrangement.[17]

Risk classification depends on the intended purpose, not simply the technology used. An AI system intended to evaluate the creditworthiness of a natural person or establish a credit score is generally listed as high-risk, subject to the Act's stated exceptions. The relevant Annex expressly excludes AI systems used for detecting financial fraud from that particular creditworthiness category. Identity verification is also not automatically high-risk: the outcome depends on whether the system performs a regulated biometric function and how it is used. Recruitment and specified biometric applications are separately addressed in Annex III.[17]

Where a system is high-risk, the provider may need to demonstrate risk management, data governance, technical documentation, logging, human-oversight design, accuracy, robustness and cybersecurity. Responsibility for data must be allocated carefully. The provider is responsible for applicable governance of training, validation and testing datasets. A deployer that controls input data is therefore required to ensure that those inputs are relevant and sufficiently representative in view of the system's intended purpose.[17] An EU bank may therefore request testing results, performance limitations, data information, logging capability, incident cooperation and change-notification rights. Such requests may reflect both the AI Act and the bank's wider obligations under EU outsourcing and digital operational-resilience rules.[27][28] They are not necessarily proof that every requested obligation applies directly to the fintech. Contracts are required to distinguish statutory requirements from the customer's internal risk standards. Changes made after deployment also matter. Rebranding a system, substantially modifying it, or changing its intended purpose may cause another actor to assume provider responsibilities. Incident arrangements are thus required to distinguish the provider's reporting duties from the deployer's obligation to monitor use, suspend a system where necessary and communicate relevant risks.[17]

Across the four countries, the EU classification analysis remains the same. The differences concern the controls already available to support it. Privacy law may assist with data mapping and impact assessment; financial-sector rules may support governance and outsourcing review; and regulatory sandboxes may provide channels for engagement. None of these removes the need for an AI-specific inventory, role analysis and evidence file.

*Scenario Two: an African bank using a European vendor*

The position is different where an African bank purchases an AI system from a European vendor. The bank will ordinarily be a deployer where it uses the system under its authority. If the system is used only in Africa and its output is not used in the EU, Article 2(1)(c) does not, on that fact alone, establish the required territorial connection. Other connections would still need to be considered.[17] The bank may nevertheless receive AI Act-related documentation, restrictions and monitoring requirements from the vendor or its multinational group. It therefore needs to obtain clear information about intended purpose, system version, performance limitations, input requirements, human oversight, logging, cybersecurity, updates and incident support. The bank should not rely solely on the vendor's classification. Classification may change if the bank deploys the system for a different purpose, integrates it into a higher-impact decision process or substantially modifies it. Equally, the customer in this case should not assume that every local configuration makes it a provider. The assessment must be based on the specific change and the criteria in Article 25.[17] Local privacy and banking rules remain independently relevant. They may require risk assessment, security, outsourcing oversight or lawful processing of personal data. Those assessments can supply useful evidence, but they do not replace AI-specific analysis of performance, human oversight, bias, system changes and EU territorial exposure.

Vendor opacity is therefore a governance issue in this circumstance. International financial-sector evidence identifies growing dependence on third-party AI systems and limited visibility into externally developed models as material risk-management concerns.[29][30] Thus, a bank that cannot obtain sufficient information should record the limitation, seek contractual remedies, restrict the use case or formally accept the residual risk through its governance process.

*Scenario Three: an African company in an EU AI supply chain*

A company providing data annotation, model testing, cloud services, software components or technical support does not automatically become an AI Act provider. Its role depends on what it supplies, the control it exercises and whether it places or modifies a system in a manner covered by the Act.

However, Article 25 requires providers of high-risk AI systems and certain third-party suppliers to specify, through written agreement, the information, technical access and assistance needed for compliance.[17] This creates a clear basis for contractual requirements even where the African supplier does not hold a principal statutory role. Contracts should address subcontracting, data provenance, security, testing methods, change notification, audit access and assistance with investigations. Intellectual-property and trade-secret protections may limit unrestricted disclosure, but they should not leave the provider unable to obtain the information required to support its own compliance. Responsibility for a defective component will also depend on the contract and other applicable law; the AI Act does not resolve every question of private liability between the parties.

*Scenario Four: a multinational group using a shared system*

A multinational group may procure one AI system centrally and use it across EU and African subsidiaries. The existence of a group-wide policy does not make the group a single actor for every purpose. Each entity's role depends on who develops, supplies, controls and uses the system. The use of African data to develop or test a system does not by itself establish that the AI Act applies. A stronger territorial connection may arise where the system is placed on the EU market or where output generated outside the EU is used by an EU group entity. Separate data-protection and cross-border transfer rules may also apply.

The practical difficulty is access to evidence. A subsidiary may be instructed to operate a centrally selected system without access to the technical file, vendor contract or change history. Group governance should therefore identify who owns the inventory, who approves the use case, who receives vendor notices, and who reports an incident. A global policy may simplify control design, but it must accommodate local financial, employment, privacy and consumer-protection law.

### *3.4 Output produced outside the EU and used inside it*

Article 2(1)(c) applies the Act to providers and deployers established in a third country where the output produced by the AI system is used in the Union.[17] This provision may capture a risk score produced in Lagos for an EU lender, a fraud alert generated in Nairobi and reviewed by an EU operations centre, a recruitment ranking produced in Johannesburg for an EU employer, or a verification result generated in Cairo and relied upon by an EU institution. The statutory text does not provide a detailed test for determining when an output is "used" in the Union. As at the cut-off date, publicly available authoritative material provides limited operational guidance on issues such as incidental access, secondary use, group-wide dashboards or outputs routed through human reviewers. These uncertainties should be stated rather than resolved through assumption.

Human review does not necessarily remove the territorial question. An output may still be used where it informs or supports a decision, although the degree of human involvement may be relevant to risk classification and other legal obligations. Organisations should therefore record where outputs are delivered, which entities can access them, the decisions they influence and whether EU use is permitted under the contract.

### *3.5 Cross-country findings*

Four principal findings stand out. First, domestic regulation provides useful foundations but not a complete AI Act readiness system. Data-protection, cybersecurity, outsourcing and governance controls can be reused, but they do not answer role classification, high-risk classification or territorial scope. Second, control of evidence is as important as the existence of policies. A regulated bank may have mature governance but limited visibility into a vendor's model. A technically capable fintech may possess detailed testing results but lack formal role analysis or contractual documentation. These are organisational possibilities, not conclusions that apply uniformly to every firm in a country or sector. Third, contractual exposure may be broader than direct statutory exposure. EU clients may request information or assurances because of their own regulatory duties. Those requests should be assessed and negotiated rather than automatically accepted as a complete statement of the supplier's legal obligations. Fourth, proportionality is essential. A common baseline should apply to every organisation: an inventory, assigned responsibility, exposure mapping, vendor review and retained evidence. More extensive validation, independent assurance and monitoring should follow where the use is high-risk, the organisation controls system development, or the consequences for individuals and markets are greater.

### *3.6 Implications for the target audiences*

For the European AI Office, the principal need is clearer practical guidance on third-country outputs, outsourced services and cross-border role allocation. Multinational companies need contracts and group policies that identify responsibilities precisely and give subsidiaries access to relevant evidence. African regulators can assist by mapping existing sectoral controls against AI-governance requirements without treating EU law as a substitute for domestic policy choices. Business associations can reduce cost and duplication through common inventories, questionnaires, contract guidance and sector-specific training.

## 4. Suggested readiness framework

This framework is designed to help an organisation determine what it knows, what it controls and what evidence it can produce. It is not a substitute for legal advice, and not every control will apply to every AI system. Its purpose is to create a defensible process for identifying direct, contractual and operational exposure. The seven steps should be applied to each AI system or use case rather than to the organisation in the abstract.

**Step 1: Map EU exposure**

The organisation should first identify every connection between its AI activities and the EU. This includes EU clients, end users, group companies, vendors and supply-chain partners. It should also identify systems offered on the EU market, outputs used in EU decisions, and contracts referring to the AI Act or EU regulatory requirements. The exercise should answer four basic questions:

- Is the system or model supplied to an EU customer?
- Can an output generated outside the EU be accessed or relied upon in the EU?
- Does an EU group company control or use the system?
- Has the organisation accepted AI-related duties through a contract?

The result should be an EU exposure map linking each system to the relevant customer, country, contract and business process.

**Step 2: Create an AI inventory**

An organisation cannot classify or govern systems it has not identified. The inventory should include systems developed internally, purchased from vendors, embedded in other software and accessed through general-purpose AI services.

For each entry, the organization should maintain an inventory as shown in Table 3. The inventory should be updated when a system, model, vendor or intended purpose changes. Inventory and context mapping are also consistent with the "Map" function of the NIST AI Risk Management Framework.[31]

| Inventory field | Information required |
|---|---|
| Purpose | The task the system performs and the decision it supports |
| Ownership | Business owner, developer, vendor and responsible function |
| Use | Users, affected persons and geographical locations |
| Data | Main inputs, outputs and personal or sensitive data involved |
| Integration | Connected systems, APIs and downstream uses |
| Significance | Whether the output determines, recommends or merely informs a decision |
| Evidence | Contracts, instructions, testing records and assessments currently available |

**Table 3: Suggested AI inventory.**

**Step 3: Classify roles and risk**

For each system, determine whether the organisation may be acting as a provider, deployer, importer, distributor, authorised representative, product manufacturer, down-stream provider, subcontractor or customer without a formal AI Act role. The organisation should then assign one preliminary risk category:

- potentially prohibited;
- potentially high-risk;
- subject to Article 50 transparency requirements;
- connected to a general-purpose AI model;
- lower-risk but contractually significant;
- apparently outside scope; or
- uncertain—specialist review required.

Classification must consider the intended purpose, actual use, affected persons and any modification made to the system. The organisation should record the reasoning and the evidence used, because legal status cannot be determined reliably from a product name or vendor description alone.[17]

**Step 4: Establish governance**

Each material AI system should have a named business owner and a documented oversight route involving legal, compliance, technology, information security and other relevant functions. Responsibilities should be proportionate to the system's risk and the organisation's role. The minimum governance questions are:

- Who approves deployment and material changes?
- Who decides whether specialist review is required?
- Who monitors the vendor and system performance?
- Who can suspend use?
- Who reports incidents and complaints?
- Which committee receives material risk information?

Staff operating or overseeing relevant systems should receive role-appropriate AI literacy and training. Article 4 requires providers and deployers to take measures, to their best extent, to ensure a sufficient level of AI literacy among relevant personnel, taking account of their knowledge, experience and context of use.[17]

**Step 5: Strengthen contracts and vendor controls**

Contracts should define the system's intended and permitted use, the parties' functions, access to documentation and the support required throughout the system's lifecycle. Priority provisions include:

- performance measures and known limitations;
- data provenance and input requirements;
- testing and validation information;
- audit and assurance rights;
- model updates and material changes;
- subcontracting;
- security and incident notification;
- regulatory cooperation;
- record retention;

- termination and transition support; and
- responsibility and remedies where obligations are not met.

A contract may allocate tasks, information rights, liability and remedies between parties. It cannot, however, determine statutory status where the facts and the AI Act assign a role differently. EU financial-sector outsourcing and operational-resilience rules also demonstrate the importance of written responsibilities, access rights, subcontracting controls and exit arrangements for important technology providers.[27][28]

**Step 6: Build and retain evidence**

Readiness should be demonstrated through records rather than policy statements alone. The evidence file for a material system may include:

- the inventory and classification record;
- legal, privacy and risk assessments;
- testing and validation results;
- instructions and model documentation;
- human-oversight procedures;
- vendor questionnaires and contracts;
- approval and decision logs;
- staff-training records;
- incident and complaint records; and
- system-change documentation.

The evidence should be proportionate to the system's function and impact. ISO/IEC 42001 similarly treats AI governance as a management system involving defined policies, responsibilities, risk processes and continual improvement.[32]

**Step 7: Monitor, respond and review**

Assessment should continue after deployment. The organisation should monitor system and model updates, performance deterioration, complaints, incidents, vendor changes, new EU customers and changes in intended purpose. A reassessment should be triggered where:

- the system begins supporting a new decision;
- an output becomes available to an EU entity;
- a vendor changes the underlying model;
- the organisation modifies or rebrands the system;
- performance or fairness concerns emerge; or
- relevant law, guidance or contractual requirements change.

Material issues should be escalated to senior management or the appropriate risk committee. A readiness maturity scale as shown in Table 4 more be created for this purpose. Scores should be assigned separately to exposure mapping, inventory, classification, governance, contracting, evidence and monitoring. A single average score may conceal a serious weakness, such as strong governance but no access to vendor documentation. Based on this, Table 5 shows the minimum action that should be taken by various kinds of organizations, based on their peculiar circumstances. The framework's central principle is hinged on the need for organisation to be able to explain what the system does, where it is used, what role it performs, who controls the necessary evidence and how risks are monitored over time.

| Level | Readiness indicator |
|---|---|
| **0 — Unaware** | No reliable inventory, owner or documented process |
| **1 — Initial** | Informal awareness and isolated controls |
| **2 — Developing** | Documented controls for selected systems |
| **3 — Operational** | Consistent governance, evidence and monitoring |
| **4 — Assured** | Independent validation, testing and continual improvement |

**Table 4: Readiness maturity scale**.

| Organisation | Minimum priority actions |
|---|---|
| **Start-up** | Identify EU connections, maintain a basic inventory, document intended use and review vendor terms |
| **Established technology supplier** | Add formal role classification, testing records, change controls and customer-assurance procedures |
| **Bank or regulated organisation** | Integrate AI review into model risk, privacy, outsourcing, cybersecurity and committee oversight |
| **Multinational subsidiary** | Clarify group roles, obtain central documentation and establish local escalation and incident responsibilities |
| **EU-facing outsourced provider** | Map outputs used in the EU, document subcontractors and retain evidence supporting customer due diligence |

**Table 5: Minimum actions by organisation type.**

## 5. Recommendations

The present analysis indicates that effective readiness will require coordinated action by European institutions, African regulators, private organisations and industry bodies. Responsibility should not rest solely on individual non-EU suppliers attempting to interpret the Act without access to complete information or specialist support. The following recommendations are there-fore suggested for the relevant actors in line with their respective roles.

### *5.1 European AI Office: publish dedicated third-country guidance*

The European AI Office needs to publish practical guidance for organisations with no establishment in the EU. Existing materials explain the Act's principal roles and obligations, but further worked examples would help clarify how these rules apply to remote service provision, outsourced operations and cross-border AI supply chains.

The guidance should address when an output produced outside Europe is considered to be used in the EU, the position of non-EU subcontractors, the function of authorised representatives and the consequences of modifying or rebranding a system. It should also explain how organisations using European vendors should assess the information provided by those vendors. Financial-services examples involving credit assessment, identity verification, fraud monitoring and customer profiling would make the guidance more practical. Evidence expectations should be proportionate to an organisation's role, size and control over the system, consistent with the Act's support measures for smaller enterprises.[17][33]

### *5.2 African regulators: develop interoperable sectoral guidance*

African financial, data-protection and technology regulators need to coordinate guidance showing how existing domestic controls can support AI governance. The objective should not be to reproduce the EU AI Act as national law. Instead, regulators should map privacy, outsourcing, cybersecurity, consumer-protection, model-risk and corporate-governance requirements against common readiness functions such as inventories, impact assessment, vendor oversight and incident management. This approach would allow organisations to reuse established controls while identifying areas where EU-facing activity requires additional analysis. It would also support the African Union Continental Artificial Intelligence Strategy's emphasis on responsible governance, institutional capacity and regional cooperation.[26]

### *5.3 EU clients and multinational companies: apply proportionate supplier requirements*

EU clients and multinational groups should avoid imposing undefined obligations re-quireing every supplier to demonstrate "full AI Act compliance." Supplier requirements should reflect the supplier's actual role, the system's intended purpose and the information genuinely needed by the regulated customer. Contracts should distinguish statutory requirements from internal group preferences. Customers should provide rele-vant system information, use proportionate audit rights and establish clear processes for changes, incidents and regulatory cooperation. Where a smaller strategic supplier has a remediable documentation or governance gap, structured support may be more effective than transferring broad and unqualified liability.

### *5.4 Non-EU organisations: adopt dual-track readiness*

African and other non-EU organisations should assess two forms of exposure separately: (i) direct legal applicability under the AI Act; and (ii) contractual, supply-chain and market-access exposure.

Uncertainty about territorial application should not delay basic readiness measures. Organisations should identify relevant AI systems, assign responsibility, map where outputs are used and review important customer and vendor contracts. The scope and cost of additional controls should then reflect the system's risk, the organisation's role and the consequences of failure.

### *5.5 Business associations: create shared implementation resources*

Banking, fintech and technology associations should develop common questionnaires, role-classification decision trees, model contract provisions, sector-specific use-case libraries and readiness assessments. Shared training, regulator engagement forums and pooled access to legal and technical expertise could reduce duplicated work, particularly for smaller organisations. These resources should remain adaptable because classification and contractual responsibilities depend on the facts of each system rather than on sector labels alone.

## 6. Limitations

It is important to note the limitations of this study, based on the purpose which it is intended to serve. First, the study relies on publicly available documents and does not include interviews, confidential contracts or internal compliance records. It may therefore not fully capture how organisations allocate AI-related responsibilities in practice. Public evidence may also overrepresent larger and more visible organisations, while the quantity and quality of available information differ across South Africa, Nigeria, Kenya and Egypt. Second, the EU AI Act remains an evolving regulatory framework. Guidance, harmonised standards, codes of practice and enforcement approaches may continue to develop after the study's legal cut-off date. Enforcement evidence is also limited immediately after August 2026, while questions of territorial applicability (particularly where outputs are produced outside the EU but used within it) remain highly dependent on the facts of each case. Third, the four countries were selected purposively and cannot represent the full legal, institutional and commercial diversity of Africa. Regulatory analogies drawn from the GDPR, financial-sector outsourcing rules and other EU regimes are also informative rather than determinative because those frameworks differ from the AI Act in scope and enforcement design.

These limitations are mitigated through source triangulation, a transparent evidence hierarchy, a fixed legal cut-off date, consistent comparative questions and scenario testing. Areas of uncertainty are identified expressly. The resulting framework should therefore be understood as a policy readiness tool, not a certification mechanism or substitute for a more detailed legal advice.

## 7. Conclusion

From the foregoing, the practical reach of the EU AI Act will also be shaped by contracts, procurement standards, multinational group controls, vendor relationships and the location in which AI outputs are ultimately used. For African organisations, readiness therefore depends on understanding not only whether the Act applies, but how regulatory expectations move through cross-border business relationships. The analysis shows that South Africa, Nigeria, Kenya and Egypt possess relevant foundations in data protection, financial regulation, cybersecurity and corporate governance. These foundations can support readiness, but they do not remove the need for AI-specific inventories, role classification, contractual review and evidence management.

A proportionate response should thus, begin with exposure mapping and proceed according to the organisation's role, the purpose of the system and the significance of its outputs. Regulators and industry bodies should support this process through clear guidance and shared tools, while EU clients should avoid imposing vague compliance obligations that suppliers cannot meaningfully interpret or demonstrate. The organisations best prepared after August 2026 will not be those that claim universal compliance, but those that can identify their exposure, explain their decisions and produce credible evidence.

**Sources**